# The asynchronous polar V1432 Aquilae
# and its path back to synchronism

*David Boyd*
*CBA (Oxford), 5 Silver Lane, West Challow, Wantage, OX12 9TX, United Kingdom*
*davidboyd@orion.me.uk*

*Joseph Patterson*
*Department of Astronomy, Columbia University, 550 West 120th Street, New York, NY 10027, USA*
*jop@astro.columbia.edu*

*William Allen*
*CBA (Blenheim), Vintage Lane Observatory, 83 Vintage Lane, RD 3, Blenheim 7273, New Zealand*

*Greg Bolt*
*CBA (Perth), 295 Camberwarra Drive, Craigie, Western Australia 6025, Australia*

*Michel Bonnardeau*
*Le Pavillon, 38930 Lalley, France*

*Tut and Jeannie Campbell*
*CBA (Arkansas), Whispering Pine Observatories, 7021 Whispering Pine Rd., Harrison, AR 72601, USA*

*David Cejudo*
*CBA (Madrid), Observatorio El Gallinero, El Berrueco, Madrid, Spain*

*Michael Cook*
*CBA (Ontario), Newcastle Observatory, 9 Laking Drive, Newcastle, Ontario, CANADA L1B 1M5*

*Enrique de Miguel*
*CBA (Huelva), Observatorio Astronomico del CIECEM, Parque Dunar Matalascañas, 21760 Almonte, Huelva, Spain*

*Claire Ding*
*Department of Astronomy, Columbia University, 550 West 120th Street, New York, NY 10027, USA*

*Shawn Dvorak*
*CBA (Orlando), 1643 Nightfall Drive, Clermont, FL 34711, USA*

*Jerrold L. Foote*
*CBA (Utah), 4175 E. Red Cliffs Drive, Kanab, UT 84741, USA*

*Robert Fried*
*Deceased, formerly at Braeside Observatory, Flagstaff, AZ 86002, USA*

*Franz-Josef Hambsch*
*CBA (Mol), Oude Bleken 12, B-2400 Mol, Belgium*

*Jonathan Kemp*
*Department of Physics, Middlebury College, Middlebury, VT 05753, USA*

*Thomas Krajci*
*CBA (New Mexico), PO Box 1351 Cloudcroft, NM 88317, USA*

*Berto Monard*
*CBA (Pretoria), PO Box 281, Calitzdorp 6661, Western Cape, South Africa*

*Yenal Ogmen*
*CBA (Cyprus), Green Island Observatory, Gecitkale, North Cyprus*

*Robert Rea*
*CBA (Nelson), Regent Lane Observatory, 8 Regent Lane, Richmond, Nelson 7020, New Zealand*

*George Roberts*
*CBA (Tennessee), 2007 Cedarmont Dr., Franklin, TN 37067, USA*

*David Skillman*
*CBA (Mountain Meadows), 6-G Ridge Road, Greenbelt, MD 20770, USA*

*Donn Starkey*
*CBA (Indiana), 2507 CR 60, Auburn, IN 46706, USA*

*Joseph Ulowetz*
*CBA (Illinois), 855 Fair Lane, Northbrook, IL 60062, USA*

*Helena Uthas*
*Department of Astronomy, Columbia University, 550 W 120th Street, New York, NY 10027, USA*

*Stan Walker*
*CBA (Waiharara), Wharemaru Observatory, P.O. Box 173, Awanui, 0451, New Zealand*

**Abstract**

V1432 Aquilae is the only known eclipsing asynchronous polar. In this respect it is unique and therefore merits our attention. We report the results of a 15-year campaign by the globally distributed Center for Backyard Astrophysics to observe V1432 Aql and investigate its return to synchronism. Originally knocked out of synchrony by a nova explosion before observing records began, the magnetic white dwarf in V1432 Aql is currently rotating slower than the orbital period but is gradually catching up. The fortuitously high inclination of the binary orbit affords us the bonus of eclipses providing a regular clock against which these temporal changes can be assessed. At the present rate, synchronism should be achieved around 2100. The continually changing trajectory of the accretion stream as it follows the magnetic field lines of the rotating white dwarf produces a complex pattern of light emission which we have measured and documented, providing comprehensive observational evidence against which physical models of the system can be tested.

## 1. Introduction

V1432 Aquilae, located at RA 19h 40m 11.42s, Dec -10° 25' 25.8" (J2000), is one of only four known asynchronous polars, the other three being V1500 Cyg, BY Cam and CD Ind. Polars are AM Her type cataclysmic variables (CVs) in which the white dwarf (WD) has a sufficiently strong magnetic field, typically >10MG, that formation of an accretion disc is inhibited. Matter transferring from the main sequence star forms an accretion stream which is diverted by the magnetic field of the WD towards one or both of its magnetic poles. Acceleration of the accretion stream onto the surface of the WD releases energy across the electromagnetic spectrum from X-rays to infrared. In most polars the interaction between the magnetic fields of the WD and the secondary results in rotation of the WD being synchronised with the binary orbital period (Cropper 1990). The common wisdom is that the departure from synchronism seen in these four systems is the result of a relatively recent nova explosion which disrupted the magnetic interaction and that these systems are in the process of slowly returning to synchronism. V1500 Cyg did indeed experience such a nova explosion in 1975. Of the four known asynchronous polars, V1432 Aql is the only one which displays orbital eclipses. These provide a regular clock against which temporal changes in the system can be assessed.

Early observations of this variable optical and X-ray source were attributed to the Seyfert galaxy NGC6814 and caused considerable excitement at the time. It was later identified as a separate source (Madejski et al. 1993) and labelled RX J1940.1-1025 before eventually being given the GCVS designation V1432 Aql.

The object was identified as a likely AM Her type CV by Staubert et al. (1994). The presence of two distinct signals in the optical light curve at 12116.6±0.5s and 12149.6±0.5s was first reported by Friedrich et al. (1994). Patterson et al. (1995) confirmed these two periods and noted that the shorter period is the binary orbital period since it is marked by deep eclipses and is consistent with radial velocity measurements. Patterson et al. also suggested that the longer period, which was consistent with a period previously observed in X-rays produced in the accretion process, was the rotation period of the WD thereby identifying V1432 Aql as an asynchronous polar. It was alternatively proposed by Watson et al. (1995) that the eclipse features were occultations of the WD by the accretion stream but subsequent analyses eventually rejected this conclusion. Further photometric and spectroscopic observations at optical and X-ray wavelengths were reported by, among others, Watson et al. (1995), Staubert et al. (2003), Mukai et al. (2003), Andronov et al. (2006) and Bonnardeau (2012).

From spectroscopic data, Watson et al. (1995) concluded that the secondary is a main sequence star of spectral type M4. Friedrich et al. (2000) Doppler-mapped the system and, in the absence of evidence of a distinct accretion stream, concluded that the accretion process is in the form of a curtain over a wide range in azimuth. Mukai et al. (2003) suggested from simple modelling that accretion takes place to both poles at all times and that accreting material may almost surround the WD.

The present analysis includes substantially more data than were available to previous analyses. In section 2 we review our new data. In sections 3, 4 and 5 we analyse the orbital, WD rotation and WD spin periods. In section 6 we review the progress towards resynchronisation and in section 7 we show how observable parameters of the system vary with rotation of the WD. Section 8 identifies some unusual behaviour yet to be explained and section 9 summarises our results.

## 2. Observations

A total of 75849 photometric observations of V1432 Aql were submitted to the Center for Backyard Astrophysics (CBA) in 312 datasets by 23 observers between 1998 and 2013. These comprised a total of 1170 hours of observation. A breakdown by year is given in Table 1.

| Year | Start date | End date | No of runs | Total time (hrs) |
|---|---|---|---|---|
| 1998 | 06 May | 23 Sep | 13 | 26 |
| 1999 | 18 May | 06 Jun | 6 | 12 |
| 2000 | 21 May | 27 Oct | 12 | 30 |
| 2002 | 26 Jul | 27 Sep | 75 | 346 |
| 2007 | 07 Jul | 20 Sep | 52 | 221 |
| 2008 | 28 Jun | 22 Jul | 2 | 6 |
| 2011 | 26 Jun | 01 Nov | 45 | 165 |
| 2012 | 07 Jun | 29 Sep | 68 | 244 |
| 2013 | 03 Jun | 11 Nov | 39 | 120 |

**Table 1. Summary of observations.**

The CBA (http://cbastro.org/) is a globally distributed network of small telescopes operated by amateur astronomers interested in cooperating in the study of variable stars. Almost all observations were obtained unfiltered to maximise time resolution and signal-to-noise and used a variety of CCD cameras and telescopes in the aperture range 0.2-m to 0.4-m. All images were bias- and dark-subtracted and flat-fielded before differential magnitudes were measured with respect to nearby comparison stars. Observers used a range of comparison stars, although most observers remained loyal to one comparison star, so alignment in magnitude between datasets had to be achieved empirically. By an iterative process, datasets from different observers which overlapped in time were aligned in magnitude and then others close in time were aligned with those. Eventually all datasets were brought into alignment in magnitude with an uncertainty estimated to be less than 0.1 magnitude, thus generating an integrated and internally consistent light curve spanning 15 years for use in our subsequent analysis. All observation times were converted to Heliocentric Julian Dates (HJD).

## 3. Orbital period

The most recent published orbital ephemeris from Bonnardeau (2012) was used to assign a provisional orbital phase to every observation. Segments of light curves between orbital phases -0.05 and +0.05 were extracted and a second order polynomial fitted to the minimum of each eclipse. Eclipses were generally symmetrical and round-bottomed and were well-fitted by a quadratic from which we could find the time and magnitude at minimum. In cases where the eclipses were significantly asymmetrical, only those points close to the minimum were used to ensure we found an accurate time of minimum. Eclipses poorly defined due to a large scatter in magnitude were rejected. An estimate was made of the average uncertainty in the time of minimum for each observer during each year by computing the root mean square residual between their times of minimum and a linear ephemeris determined for all measured eclipses in that year. This estimated uncertainty was then assigned to all observations by that observer in that year. The process was iterated to test its stability. Orbital cycle numbers were assigned to each eclipse based on the mid-eclipse ephemeris in Patterson et al. (1995).

At this stage we cannot say that these times represent the true times of mid-eclipse of the WD by the secondary star since there may be sources of light external to the WD which also experience eclipse and could alter the shape of the light curve around the true time of mid-eclipse. However, for the following analysis of the orbital period, we assume that these other effects average out over time and we take our derived times of minimum as times of mid-eclipse. Orbital cycle numbers, mid-eclipse times and assigned uncertainties for 228 measured eclipses are listed in Table 2.

A further 121 eclipse times including optical and X-ray observations were extracted from published papers by Patterson et al. (1995), Watson et al. (1995), Mukai et al. (2003), Andronov et al. (2006) and Bonnardeau (2012). In some cases it was clear that the published uncertainties on these times bore little relation to their intrinsic scatter and therefore could not be used to compute reliable weights on these times. For this reason, uncertainties were recomputed as described above for the times published in each paper.

We computed the following orbital ephemeris as a function of the orbital cycle number E by a weighted linear regression using all 349 times of mid-eclipse. Each mid-eclipse time was weighted by the inverse square of its assigned uncertainty.

HJD (mid-eclipse) = 2449199.69307(6)
     + 0.140234751(3).E     (1)

We adopt the orbital period $P_{orb}$ = 0.140234751d (12116.282s) in our subsequent analysis. Our combined 15-year light curve phased on this orbital period and averaged in 100 bins is shown in Figure 1.

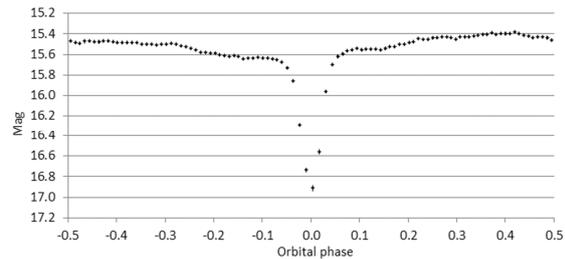

**Figure 1. Combined 15-year light curve phased on the orbital period 0.140234751d and averaged in 100 bins.**

The O-C (Observed minus Calculated) residuals to this linear ephemeris are shown in Figure 2. The rms scatter is 110s and the reduced chi-squared of this fit is 1.02.

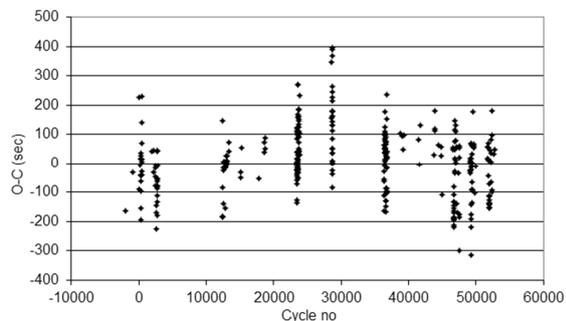

**Figure 2. O-C residuals of eclipse times to the linear orbital ephemeris in equation (1).**

The weighted eclipse times were also fitted with a 2nd order polynomial which gave the following quadratic orbital ephemeris.

HJD (mid-eclipse) = 2449199.69257(12)
     + 0.140234804(12).E
     - 9.7(2.1)x10$^{-13}$.E$^2$     (2)

The O-C residuals to this quadratic ephemeris are shown in Figure 3. This ephemeris represents a rate of change of orbital period $dP_{orb}/dt$ = -1.38(29)x10$^{-11}$ years/year. The rms scatter is 105s and the reduced chi-squared is 0.96.

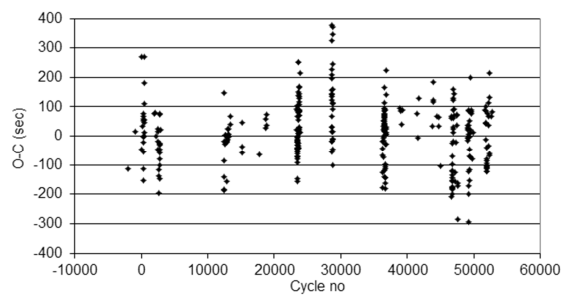

**Figure 3. O-C residuals of eclipse times to the quadratic orbital ephemeris in equation (2).**

## 4. WD rotation period

Previous analyses have concluded that the WD is slowly rotating in the rest frame of the binary system about an axis perpendicular to the orbital plane with a period which is currently about 62 days. This period is slowly lengthening due to interaction between the magnetic field of the WD and the accretion stream from the secondary which is creating a torque slowing the WD rotation. In time the WD will stop rotating relative to the secondary and the system will have resynchronised. As seen from our vantage point (which orbits around the rest frame of the binary system every 3hr 22min) the WD will then appear to spin in synchrony with the orbital period of the binary.

The WD rotation period in the rest frame of the binary, $P_{rot}$, is given by the beat period between the orbital period $P_{orb}$ and apparent WD spin period $P_{spin}$ as seen from our orbiting vantage point.

$$1/P_{rot} = 1/P_{orb} - 1/P_{spin}$$

To predict when resynchronisation will occur we need to measure $P_{spin}$ and hence find $P_{rot}$ and the rate at which it is changing.

## 5. WD spin period

Geckeler & Staubert (1997) and Staubert et al. (2003) proposed an accretion scenario based on a dipole model of the WD magnetic field. This predicts modulation in the out-of-eclipse light curve due to the changing aspect of the accretion spot on the surface of the rotating WD. To search for that signal, we first removed the eclipses in the light curve between orbital phases -0.05 and +0.05. We then carried out a period analysis of the remaining out-of-eclipse light curve year by year. This gave initial estimates of the WD spin period for each year. The uncertainty on these estimates was too large to accurately define a rate of change but they were sufficient to phase the out-of-eclipse light curve for each year on this spin period. As noted by several authors, there is a prominent 'spin dip' or 'trough' in the spin-phased light curve for each year. We took the average position of the spin dip for each year as defining spin phase 0.0 in that year. Staubert et al. (2003) suggested that this dip occurs due to absorption of light by the accretion column. The timing of these dips potentially provides a way of measuring the WD spin period accurately. We note in passing that there is possible confusion in nomenclature as some papers refer to the eclipse as a 'dip'.

Figure 4 shows the combined 15-year out-of-eclipse light curve phased within each year on the WD spin period for that year and averaged in 100 bins. We performed a period analysis on the whole 15 year out-of-eclipse dataset. As might be expected given the variation in spin period over this interval, the power spectrum contained many signals of similar strength spanning a broad range of spin period and provided little useful information. No signal was detectable at the orbital period.

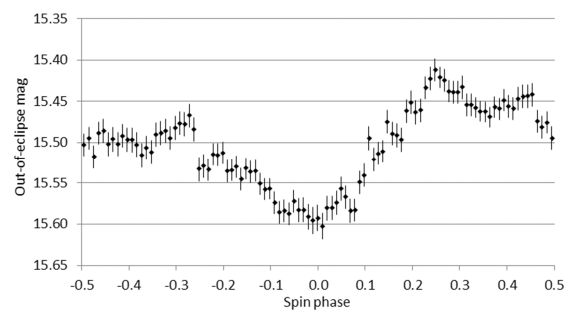

**Figure 4. Combined 15-year out-of-eclipse light curve phased within each year on the WD spin period for that year and averaged in 100 bins.**

The times of minimum of all sufficiently well-defined dips within the spin phase range -0.1 to +0.1 were measured and uncertainties assigned as described above for eclipses. A total of 172 spin dip times were measured and are listed in Table 3. A further 73 spin dip times were obtained from the published literature and uncertainties on these reassigned as before. A preliminary linear ephemeris derived from these 245 times was used to assign an initial cycle number to each spin dip. The resulting O-C residuals revealed a strongly quadratic behaviour but also evidence that our initial cycle number assignment was not perfect. The initial cycle number assignment was adjusted until we achieved smoothly-varying and internally consistent O-C residuals. This process was repeated several times with different initial linear ephemerides to check that the set of assigned cycle numbers was stable and, we therefore assume, correct. These spin cycle numbers are also listed in Table 3. The O-C residuals to a linear ephemeris are shown in Figure 5(upper). We then calculated a quadratic ephemeris whose O-C residuals are shown in Figure 5(middle). Clearly this ephemeris, while an approximate match to the data, is still not a particularly good fit. It does however provide an average value for the rate of change of the WD spin period $dP_{spin}/dt = -1.042(1) \times 10^{-8}$ years/year. This is consistent with values found by Staubert et al.

(2003) and Mukai et al. (2003) among others. By increasing the order of the ephemeris, we found that the following fifth order polynomial in the spin cycle number E provided the best fit to the data and that this could not be improved with a higher order fit.

$$\text{HJD (spin dip)} = \alpha + \beta.E + \gamma.E^2 + \delta.E^3 + \varepsilon.E^4 + \zeta.E^5 \quad (3)$$

where $\alpha = 2448921.5330(23)$
$\beta = 0.14063204(75)$
$\gamma = -2.09(76) \times 10^{-10}$
$\delta = -2.83(33) \times 10^{-14}$
$\varepsilon = 5.65(64) \times 10^{-19}$
$\zeta = -3.83(45) \times 10^{-24}$

The O-C residuals to this fifth order ephemeris are shown in Figure 5(lower). We are not claiming any physical justification for this degree of fit, simply that it provides the closest polynomial fit to the data and therefore the best basis for determining the varying WD spin and rotation periods. The reduced chi-squared for this fit was 1.00, more than a factor of two better than any lower order fit.

We also investigated an exponential fit of the form a.exp(b.x) to the spin dip times. This provided a poor fit, quite close to the initial linear ephemeris above.

Equation (3) gives the times of spin dips as a polynomial in the spin cycle number E. By differentiating this polynomial with respect to E, we obtained the following expression for the spin period as a function of E.

$$P_{spin} = \beta + 2.\gamma.E + 3.\delta.E^2 + 4.\varepsilon.E^3 + 5.\zeta.E^4 \quad (4)$$

We calculated the value of $P_{spin}$ at each spin dip time and fitted these with a fifth order polynomial in HJD. This provided a means to compute $P_{spin}$ and hence $P_{rot}$ at any time. We chose HJD = 2449544.81070, the mid-eclipse time closest to a spin dip minimum, as defining WD rotation phase 0.0. By numerical integration in 1 day steps, we determined the WD rotation phase at the time of every observation, eclipse and spin dip. The integration step size was determined empirically by reducing the step size until the computed WD rotation phase at the end of 15 years remained stable.

## 6. Progress towards resynchronization

We found that $P_{spin}$ reduced from 0.1406320d (12150.61s) in October 1992 to 0.1405542d (12143.88s) in September 2013. The corresponding increase in $P_{rot}$ was from 49.641d to 61.703d. Figure 6 illustrates these changes.

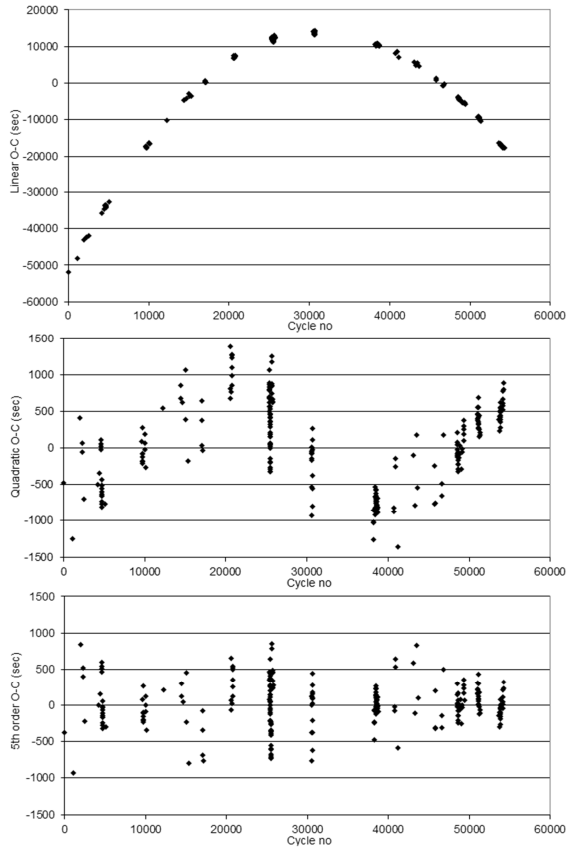

**Figure 5.** O-C residuals of spin dip times to a linear ephemeris (upper), to a quadratic ephemeris (middle), and to the best fit ephemeris in equation (3) (lower).

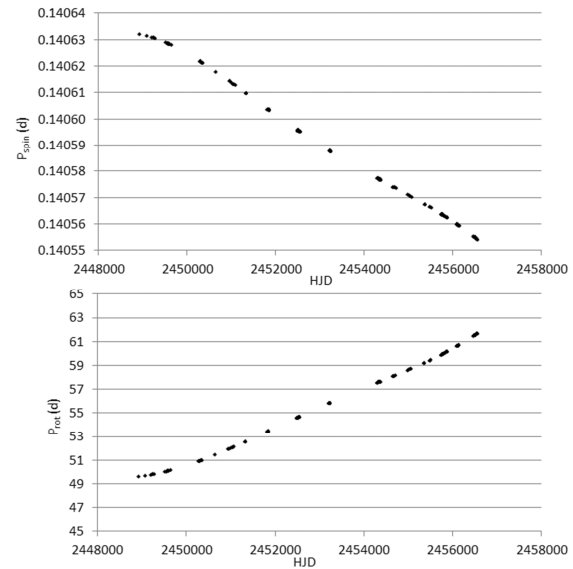

**Figure 6.** Time variation of $P_{spin}$ (upper) and $P_{rot}$ (lower).

During this interval the WD rotated 139 times in the binary rest frame. These results are consistent with values of $P_{spin}$ found at various times by Patterson et al. (1995), Geckeler & Staubert (1997), Staubert et al. (2003) and Andronov et al. (2006).

In the other three asynchronous polars the apparent WD spin period $P_{spin}$ is shorter than the orbital period $P_{orb}$. This is generally assumed to be because the WD has been spun up during a recent nova explosion and is subsequently slowing back into synchrony. Why the WD spins slower in V1432 Aql is an unanswered question.

We can extrapolate the trend in $P_{rot}$ between 1992 and 2013 to find when resynchronisation is likely to occur. This will happen when $1/P_{rot}$ becomes zero. Using all the currently available data, linear extrapolation predicts resynchronisation will occur in 2097 while quadratic extrapolation predicts it will occur in 2106. This is of the same order as the resynchronisation timescale of ~170 years found for V1500 Cyg (Schmidt et al. 1995).

## 7. Variation with WD rotation phase

We investigated how various observable properties of the system vary with the WD rotation phase in the rest frame of the binary. This provides observational evidence against which physical models of the system may be tested.

### 7.1 Variation of orbital light curve with WD rotation phase

In Figure 7 we show the averaged orbital-phased light curve in each of 10 bins of WD rotation phase. The data plotted in this and subsequent similar plots are the mean values for each bin and the error bars represent the standard error for each bin. The light curves show a broad minimum which moves forward in orbital phase as the WD rotation phase increases. Plotting the orbital phase of this minimum against the corresponding WD rotation phase (Figure 8) shows a clear linear relationship. To investigate this further we removed the eclipses and plotted the averaged out-of-eclipse light curve against the difference between the orbital and WD rotation phases (Figure 9). This highlights the effect and shows there is a drop of 0.7 magnitudes, corresponding to a flux drop of about 50%, when the orbital and the WD rotation phases are the same.

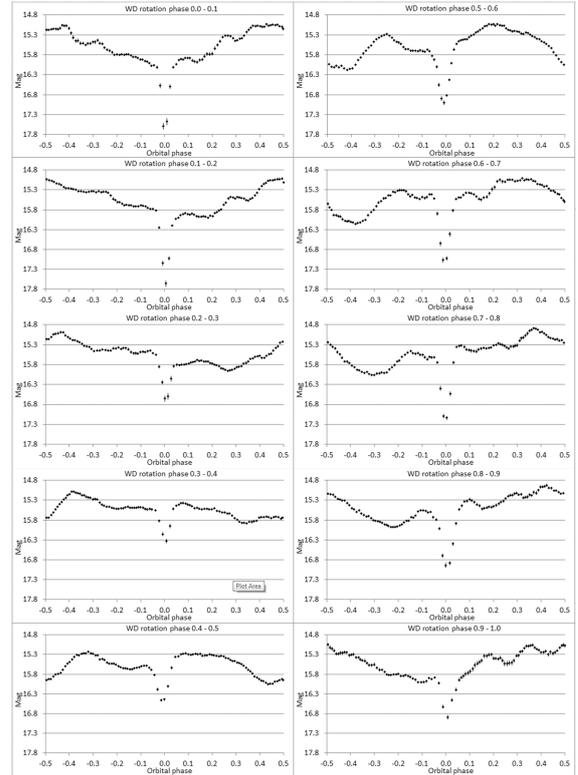

**Figure 7.** Averaged orbital-phased light curve in 10 bins of WD rotation phase.

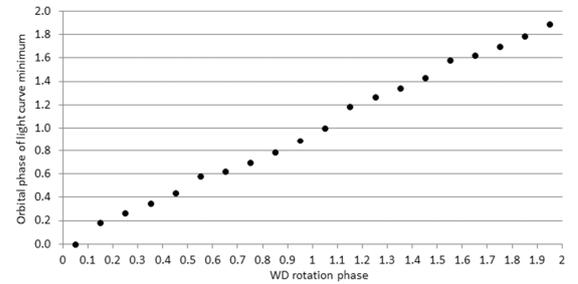

**Figure 8.** Variation of orbital phase of out-of-eclipse light curve minimum with WD rotation phase.

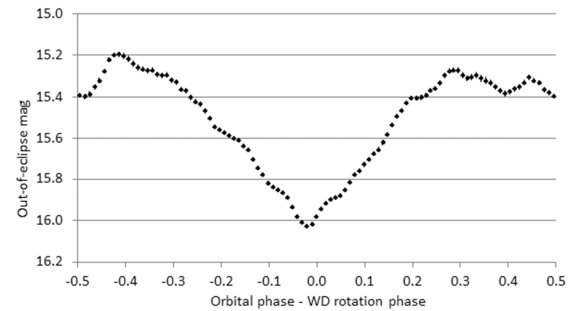

**Figure 9.** Variation of averaged out-of-eclipse light curve with the difference between orbital and WD rotation phases.

## 7.2 Variation of eclipses with WD rotation phase

Figure 10 shows how the O-C residuals of mid-eclipse times to the linear ephemeris in equation (1) vary with WD rotation phase. Variation with the quadratic ephemeris in equation (2) is virtually identical. Variation of the averaged eclipse profile with WD rotation phase is shown in Figure 11. Eclipses in the visual waveband are generally round-bottomed so either they are grazing or an extended light source is being eclipsed or both. This is in marked contrast to the steep-sided total eclipses observed in the hard X-rays originating in the relatively compact accretion regions close to the magnetic poles of the WD (Mukai et al. 2003). However we note that we find the O-C timing residuals of optical and X-ray eclipses to be fully consistent.

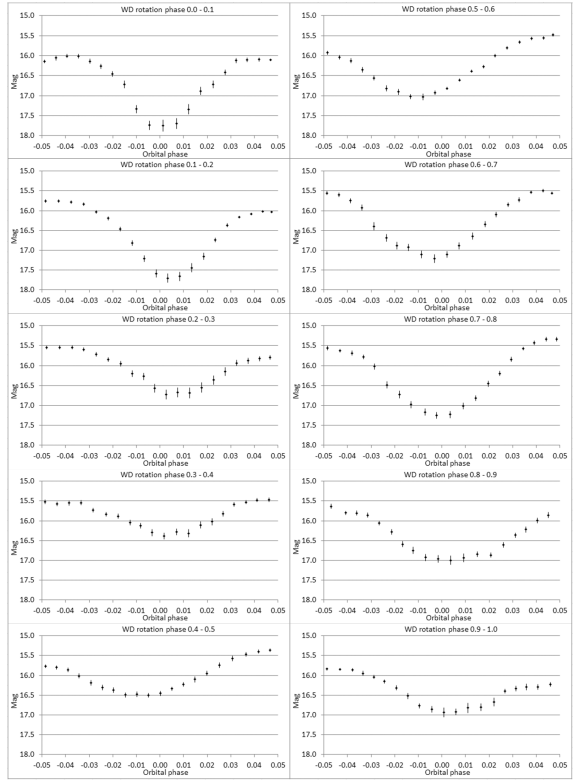

**Figure 11. Averaged eclipse profile in 10 bins of WD rotation phase.**

As well as variation in mid-eclipse time with WD rotation phase there are also noticeable variations in eclipse depth and width. To examine these variations, we first calculated the mean eclipse depth for each of 10 bins of WD rotation phase by averaging the magnitudes on either side of each eclipse in Figure 7 and subtracting this from the average magnitude at eclipse minimum calculated from the previous quadratic fits to the eclipses in that bin. We then used a quadratic fit to the lower half of each eclipse in Figure 7 to calculate the mean eclipse width at half-depth for each of 10 bins of WD rotation phase. In Figure 12 we show how eclipse depth, magnitude at eclipse minimum and eclipse width vary with WD rotation phase.

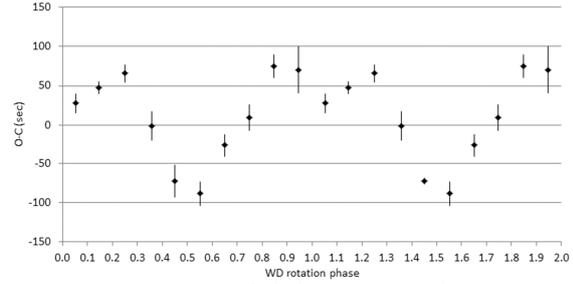

**Figure 10. Variation of O-C residuals of mid-eclipse times to the linear eclipse ephemeris in equation (1) with WD rotation phase.**

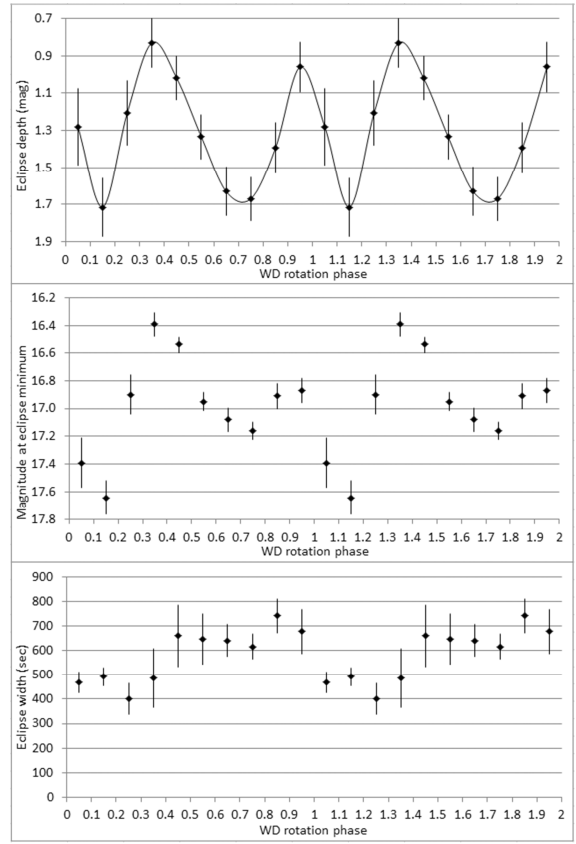

**Figure 12. Variation of eclipse depth (upper), magnitude at eclipse minimum (middle) and eclipse width (lower) with WD rotation phase (the curve is purely to help guide the eye).**

Eclipse depth shows two cycles of variation per WD rotation with maximum amplitude 0.9 magnitudes corresponding to a flux drop of 56%. Eclipse width shows one cycle of variation per WD rotation with a mean width of 463s in the phase range 0.0 to 0.4 and 664s in the range 0.4 to 1.0. This is consistent with the average width of 632±35s given in Patterson et al. (1995). Eclipse widths measured in X-rays are generally larger than those measured optically. Figure 13 shows durations of 18 X-ray eclipses given in Mukai et al. (2003) for which we have computed the WD rotation phase. There is a degree of correlation in that both optical and X-ray eclipse widths are larger in the phase range 0.8 to 1.0. Note that WD rotation phase 0.0 adopted by Mukai et al. corresponds to phase 0.42 in our analysis.

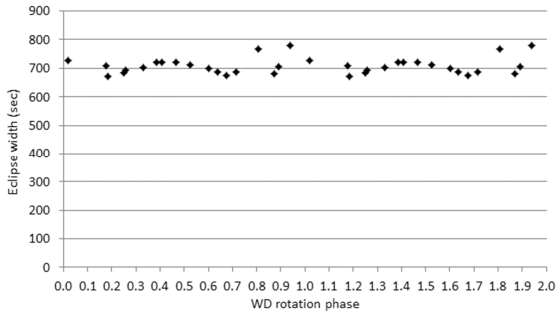

**Figure 13. Variation of X-ray eclipse width from Mukai et al. (2003) with WD rotation phase.**

### 7.3 Variation of WD spin dip with WD rotation phase

Figure 14 shows how the O-C residuals of spin dip times to the best fit ephemeris in equation (3) vary with WD rotation phase. Staubert et al. (2003) considered this variation to be a consequence of both the changing trajectory of the accretion stream in the magnetic field of the rotating WD and movement of its impact point on the WD surface. Figure 15 shows the averaged spin-phased out-of-eclipse light curve in each of 10 bins of WD rotation phase. The most noticeable feature is a persistent minimum around spin phase 0.0 between WD rotation phases 0.4 and 0.9.

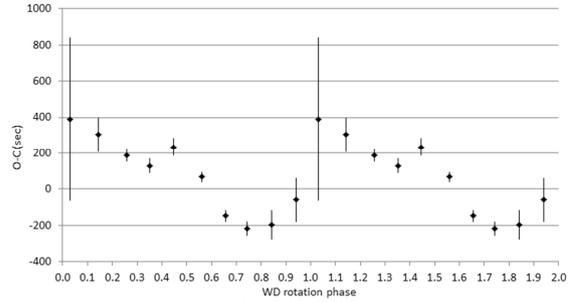

**Figure 14. Variation of O-C residuals of spin dip times to the best fit spin dip ephemeris in equation (3) with WD rotation phase.**

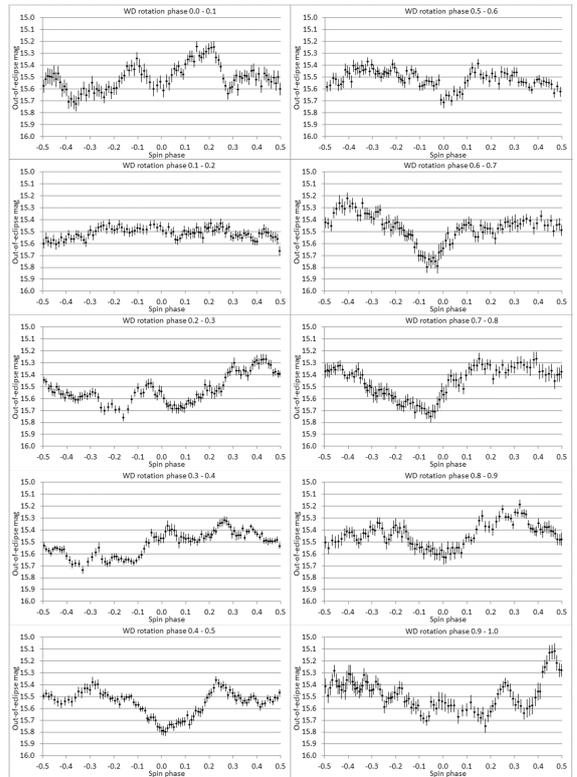

**Figure 15. Averaged spin-phased out-of-eclipse light curve in 10 bins of WD rotation phase.**

## 8. Strange behaviour in 2002

The O-C residuals of spin dip times exhibit a strange discontinuity during 2002 which is shown in Figure 16.

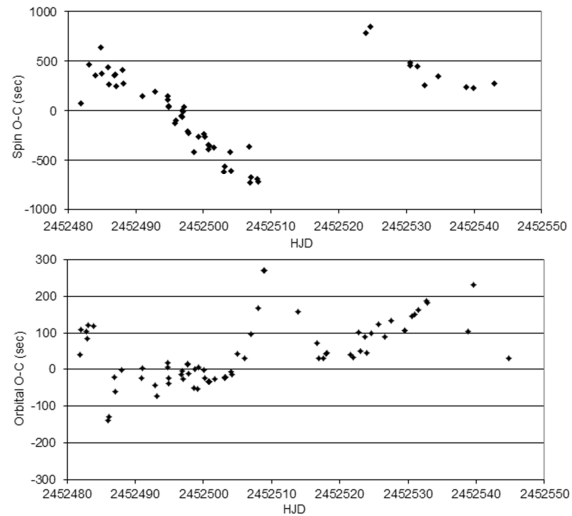

**Figure 16. Spin dip and orbital eclipse O-C residuals in 2002.**

There is also a short-lived increase in the orbital eclipse O-C residuals around the same time. The gap in the spin dip residuals is due to the temporary coincidence of the spin dip with the orbital eclipse which prevents measurement of spin dip timings. These plots include observations from several observers so this is a real effect. On four other occasions during observing seasons in 2007, 2011 (twice) and 2013 there is a coincidence of spin dip and orbital eclipse and thus a gap in the data but in none of these is there a discontinuity in the spin dip O-C residuals like the one in 2002. At the moment we do not have an explanation for this behaviour.

## 9. Summary

Our analysis of 21 years of data on the asynchronous polar V1432 Aql has revealed the following:
a) the orbital period of the binary system is 0.140234751d (12116.282s);
b) a quadratic ephemeris with $dP_{orb}/dt = -1.38 \times 10^{-11}$ years/year is slightly favoured over a linear ephemeris;
c) the apparent WD spin period has reduced from 0.1406320d (12150.61s) in October 1992 to 0.1405542d (12143.88s) in September 2013;
d) the average rate of change of the WD spin period is $-1.042 \times 10^{-8}$ years/year;
e) the WD rotation period has increased from 49.641d to 61.703d over the same interval;
f) resynchronisation is expected to occur somewhere between 2097 and 2106;
g) there is a dependency of several observable quantities on the WD rotation phase, specifically:
   • orbital-phased light curves;
   • mid-eclipse times;
   • eclipse depth;
   • eclipse width;
   • spin-phased out-of-eclipse light curves;
   • spin dip times.

## 10. Acknowledgements

D.B. gratefully acknowledges the contribution of Colin Littlefield whose ongoing work on V1432 Aql led to his undertaking this analysis. We thank the organisers of the Society for Astronomical Sciences Symposia for providing a continuing opportunity for us to present the results of pro-am collaboration.

| Orbital cycle | Mid-eclipse time (HJD) | Assigned uncertainty (d) | Orbital cycle | Mid-eclipse time (HJD) | Assigned uncertainty (d) |
|---|---|---|---|---|---|
| 12424 | 2450941.97131 | 0.00170 | 23653 | 2452516.66647 | 0.00080 |
| 12445 | 2450944.91238 | 0.00170 | 23655 | 2452516.94642 | 0.00120 |
| 12452 | 2450945.89407 | 0.00170 | 23660 | 2452517.64760 | 0.00080 |
| 12459 | 2450946.87684 | 0.00170 | 23662 | 2452517.92821 | 0.00120 |
| 12495 | 2450951.92467 | 0.00170 | 23663 | 2452518.06848 | 0.00120 |
| 12823 | 2450997.92149 | 0.00170 | 23688 | 2452521.57430 | 0.00080 |
| 13370 | 2451074.63250 | 0.00170 | 23691 | 2452521.99492 | 0.00120 |
| 13413 | 2451080.66227 | 0.00170 | 23696 | 2452522.69689 | 0.00120 |
| 15119 | 2451319.90191 | 0.00170 | 23698 | 2452522.97676 | 0.00120 |
| 15133 | 2451321.86496 | 0.00170 | 23703 | 2452523.67840 | 0.00080 |
| 15197 | 2451330.84118 | 0.00170 | 23705 | 2452523.95835 | 0.00120 |
| 17736 | 2451686.89599 | 0.00170 | 23710 | 2452524.66017 | 0.00080 |
| 18640 | 2451813.66965 | 0.00120 | 23717 | 2452525.64210 | 0.00080 |
| 18649 | 2451814.93139 | 0.00060 | 23724 | 2452526.62333 | 0.00110 |
| 18784 | 2451833.86321 | 0.00120 | 23731 | 2452527.60549 | 0.00110 |
| 18806 | 2451836.94880 | 0.00060 | 23745 | 2452529.56847 | 0.00110 |
| 23405 | 2452481.88785 | 0.00120 | 23745 | 2452529.56847 | 0.00080 |
| 23406 | 2452482.02892 | 0.00120 | 23752 | 2452530.55056 | 0.00110 |
| 23412 | 2452482.87027 | 0.00120 | 23755 | 2452530.97133 | 0.00120 |
| 23413 | 2452483.01027 | 0.00120 | 23759 | 2452531.53239 | 0.00110 |
| 23414 | 2452483.15093 | 0.00120 | 23767 | 2452532.65455 | 0.00120 |
| 23419 | 2452483.85208 | 0.00120 | 23768 | 2452532.79474 | 0.00120 |
| 23434 | 2452485.95261 | 0.00120 | 23811 | 2452538.82392 | 0.00120 |
| 23435 | 2452486.09296 | 0.00120 | 23817 | 2452539.66681 | 0.00120 |
| 23441 | 2452486.93560 | 0.00120 | 23854 | 2452544.85315 | 0.00060 |
| 23442 | 2452487.07539 | 0.00120 | 36376 | 2454300.87311 | 0.00110 |
| 23449 | 2452488.05770 | 0.00120 | 36472 | 2454314.33505 | 0.00120 |
| 23469 | 2452490.86214 | 0.00120 | 36479 | 2454315.31708 | 0.00120 |
| 23470 | 2452491.00269 | 0.00120 | 36480 | 2454315.45684 | 0.00120 |
| 23484 | 2452492.96545 | 0.00120 | 36486 | 2454316.29815 | 0.00120 |
| 23485 | 2452493.10534 | 0.00120 | 36487 | 2454316.43881 | 0.00120 |
| 23497 | 2452494.78907 | 0.00050 | 36493 | 2454317.28026 | 0.00120 |
| 23497 | 2452494.78920 | 0.00120 | 36507 | 2454319.24369 | 0.00120 |
| 23498 | 2452494.92880 | 0.00120 | 36508 | 2454319.38380 | 0.00120 |
| 23498 | 2452494.92895 | 0.00050 | 36514 | 2454320.22538 | 0.00120 |
| 23511 | 2452496.75214 | 0.00050 | 36521 | 2454321.20711 | 0.00120 |
| 23512 | 2452496.89248 | 0.00050 | 36522 | 2454321.34760 | 0.00120 |
| 23513 | 2452497.03245 | 0.00120 | 36529 | 2454322.32891 | 0.00120 |
| 23518 | 2452497.73407 | 0.00120 | 36530 | 2454322.46942 | 0.00120 |
| 23518 | 2452497.73409 | 0.00050 | 36536 | 2454323.31105 | 0.00120 |
| 23519 | 2452497.87402 | 0.00050 | 36543 | 2454324.29277 | 0.00120 |
| 23524 | 2452498.57475 | 0.00110 | 36544 | 2454324.43305 | 0.00120 |
| 23525 | 2452498.71557 | 0.00110 | 36550 | 2454325.27433 | 0.00120 |
| 23528 | 2452499.13566 | 0.00060 | 36551 | 2454325.41490 | 0.00120 |
| 23529 | 2452499.27658 | 0.00060 | 36557 | 2454326.25544 | 0.00120 |
| 23535 | 2452500.11789 | 0.00060 | 36558 | 2454326.39607 | 0.00120 |
| 23536 | 2452500.25787 | 0.00060 | 36561 | 2454326.81610 | 0.00110 |
| 23539 | 2452500.67848 | 0.00050 | 36564 | 2454327.23696 | 0.00120 |
| 23539 | 2452500.67848 | 0.00120 | 36586 | 2454330.32243 | 0.00120 |
| 23540 | 2452500.81869 | 0.00120 | 36587 | 2454330.46293 | 0.00120 |
| 23540 | 2452500.81871 | 0.00050 | 36593 | 2454331.30414 | 0.00120 |
| 23546 | 2452501.66021 | 0.00110 | 36594 | 2454331.44441 | 0.00120 |
| 23556 | 2452503.06257 | 0.00120 | 36607 | 2454333.26534 | 0.00120 |
| 23556 | 2452503.06261 | 0.00060 | 36611 | 2454333.82602 | 0.00110 |
| 23557 | 2452503.20285 | 0.00060 | 36621 | 2454335.23023 | 0.00120 |
| 23563 | 2452504.04442 | 0.00060 | 36629 | 2454336.35290 | 0.00120 |
| 23564 | 2452504.18457 | 0.00060 | 36664 | 2454341.25801 | 0.00120 |
| 23570 | 2452505.02661 | 0.00120 | 36665 | 2454341.39846 | 0.00120 |
| 23577 | 2452506.00811 | 0.00120 | 36672 | 2454342.38032 | 0.00120 |
| 23584 | 2452506.99055 | 0.00120 | 36679 | 2454343.36197 | 0.00120 |
| 23591 | 2452507.97303 | 0.00120 | 36686 | 2454344.34412 | 0.00120 |
| 23597 | 2452508.81561 | 0.00120 | 36700 | 2454346.30739 | 0.00120 |
| 23598 | 2452508.95584 | 0.00120 | 36701 | 2454346.44747 | 0.00120 |
| 23633 | 2452513.86275 | 0.00120 | 36736 | 2454351.35576 | 0.00120 |

| Orbital cycle | Mid-eclipse time (HJD) | Assigned uncertainty (d) |
|---|---|---|
| 36774 | 2454356.68560 | 0.00110 |
| 36781 | 2454357.66765 | 0.00110 |
| 36793 | 2454359.35073 | 0.00120 |
| 36795 | 2454359.63099 | 0.00110 |
| 36802 | 2454360.61279 | 0.00110 |
| 36817 | 2454362.71860 | 0.00110 |
| 36828 | 2454364.25867 | 0.00120 |
| 46631 | 2455738.98035 | 0.00120 |
| 46638 | 2455739.96194 | 0.00120 |
| 46653 | 2455742.06235 | 0.00120 |
| 46706 | 2455749.49564 | 0.00120 |
| 46706 | 2455749.49577 | 0.00140 |
| 46707 | 2455749.63560 | 0.00120 |
| 46719 | 2455751.31885 | 0.00120 |
| 46720 | 2455751.45882 | 0.00120 |
| 46720 | 2455751.45966 | 0.00140 |
| 46721 | 2455751.59893 | 0.00120 |
| 46727 | 2455752.44007 | 0.00120 |
| 46728 | 2455752.58004 | 0.00120 |
| 46728 | 2455752.58025 | 0.00140 |
| 46735 | 2455753.56188 | 0.00120 |
| 46748 | 2455755.38592 | 0.00120 |
| 46749 | 2455755.52527 | 0.00120 |
| 46756 | 2455756.50686 | 0.00120 |
| 46836 | 2455767.72743 | 0.00110 |
| 46857 | 2455770.67299 | 0.00110 |
| 46858 | 2455770.81266 | 0.00110 |
| 46865 | 2455771.79521 | 0.00110 |
| 46871 | 2455772.63696 | 0.00110 |
| 46872 | 2455772.77764 | 0.00110 |
| 46886 | 2455774.74127 | 0.00110 |
| 46936 | 2455781.75133 | 0.00110 |
| 46940 | 2455782.31240 | 0.00120 |
| 47228 | 2455822.69791 | 0.00170 |
| 47341 | 2455838.54598 | 0.00100 |
| 47461 | 2455855.37520 | 0.00100 |
| 47484 | 2455858.60066 | 0.00200 |
| 47489 | 2455859.30139 | 0.00100 |
| 47527 | 2455864.62802 | 0.00200 |
| 47534 | 2455865.60823 | 0.00200 |
| 47541 | 2455866.59119 | 0.00200 |
| 49127 | 2456089.00534 | 0.00120 |
| 49128 | 2456089.14559 | 0.00120 |
| 49134 | 2456089.98689 | 0.00120 |
| 49136 | 2456090.26786 | 0.00120 |
| 49143 | 2456091.24949 | 0.00120 |
| 49149 | 2456092.09085 | 0.00120 |
| 49150 | 2456092.23124 | 0.00120 |
| 49227 | 2456103.02993 | 0.00120 |
| 49234 | 2456104.01062 | 0.00120 |
| 49247 | 2456105.83402 | 0.00330 |
| 49249 | 2456106.11507 | 0.00120 |
| 49264 | 2456108.21419 | 0.00120 |
| 49276 | 2456109.89809 | 0.00330 |
| 49277 | 2456110.03923 | 0.00120 |
| 49278 | 2456110.17889 | 0.00120 |
| 49311 | 2456114.80858 | 0.00330 |
| 49325 | 2456116.77011 | 0.00330 |
| 49368 | 2456122.80118 | 0.00330 |
| 49397 | 2456126.86745 | 0.00330 |
| 49425 | 2456130.79598 | 0.00330 |
| 49430 | 2456131.49602 | 0.00110 |
| 49437 | 2456132.47844 | 0.00110 |
| 49444 | 2456133.46078 | 0.00110 |
| 49451 | 2456134.44161 | 0.00110 |
| 49468 | 2456136.82462 | 0.00330 |
| 49480 | 2456138.51058 | 0.00110 |
| 49615 | 2456157.44091 | 0.00110 |
| 49629 | 2456159.40428 | 0.00110 |
| 49636 | 2456160.38579 | 0.00110 |
| 49757 | 2456177.35238 | 0.00110 |
| 49921 | 2456200.35192 | 0.00110 |
| 51741 | 2456455.57996 | 0.00100 |
| 51755 | 2456457.54278 | 0.00100 |
| 51764 | 2456458.80419 | 0.00110 |
| 51791 | 2456462.59118 | 0.00100 |
| 51798 | 2456463.57281 | 0.00100 |
| 51876 | 2456474.51163 | 0.00100 |
| 51877 | 2456474.65204 | 0.00100 |
| 51940 | 2456483.48445 | 0.00100 |
| 51941 | 2456483.62459 | 0.00100 |
| 51948 | 2456484.60660 | 0.00100 |
| 51949 | 2456484.74633 | 0.00100 |
| 51950 | 2456484.88675 | 0.00100 |
| 51956 | 2456485.72894 | 0.00100 |
| 51963 | 2456486.70965 | 0.00100 |
| 51964 | 2456486.85030 | 0.00100 |
| 51970 | 2456487.69159 | 0.00100 |
| 51978 | 2456488.81329 | 0.00110 |
| 52134 | 2456510.69206 | 0.00110 |
| 52261 | 2456528.50230 | 0.00100 |
| 52269 | 2456529.62250 | 0.00110 |
| 52276 | 2456530.60530 | 0.00110 |
| 52277 | 2456530.74518 | 0.00110 |
| 52339 | 2456539.43858 | 0.00100 |
| 52346 | 2456540.42017 | 0.00100 |
| 52383 | 2456545.61210 | 0.00220 |
| 52412 | 2456549.67795 | 0.00220 |
| 52583 | 2456573.65734 | 0.00110 |
| 52746 | 2456596.51577 | 0.00110 |

**Table 2.** Orbital cycle numbers, mid-eclipse times and assigned uncertainties.

| Spin cycle | Spin dip time (HJD) | Assigned uncertainty (d) |
|---|---|---|
| 0 | 2450944.98512 | 0.00520 |
| 7 | 2450945.96734 | 0.00520 |
| 638 | 2451034.69852 | 0.00520 |
| 922 | 2451074.61808 | 0.00520 |
| 2652 | 2451317.88399 | 0.00520 |
| 2666 | 2451319.84949 | 0.00520 |
| 2744 | 2451330.81220 | 0.00520 |
| 6178 | 2451813.66523 | 0.00260 |
| 6187 | 2451814.92921 | 0.00260 |
| 6343 | 2451836.87035 | 0.00260 |
| 6357 | 2451838.83669 | 0.00260 |
| 6272 | 2451826.88156 | 0.00260 |
| 6329 | 2451834.89853 | 0.00260 |
| 6336 | 2451835.88560 | 0.00260 |
| 6350 | 2451837.85438 | 0.00260 |
| 6400 | 2451844.87992 | 0.00260 |
| 10939 | 2452483.06526 | 0.00470 |
| 10946 | 2452484.04821 | 0.00470 |
| 10952 | 2452484.89503 | 0.00470 |
| 10953 | 2452485.03256 | 0.00470 |
| 10959 | 2452485.87689 | 0.00470 |
| 10960 | 2452486.01551 | 0.00470 |
| 10966 | 2452486.86014 | 0.00470 |
| 10967 | 2452487.00087 | 0.00470 |
| 10968 | 2452487.14011 | 0.00470 |
| 10974 | 2452487.98550 | 0.00470 |
| 10975 | 2452488.12455 | 0.00470 |
| 10995 | 2452490.93498 | 0.00470 |
| 11009 | 2452492.90387 | 0.00470 |
| 11022 | 2452494.73064 | 0.00470 |
| 11022 | 2452494.73102 | 0.00470 |
| 11023 | 2452494.87045 | 0.00470 |
| 11023 | 2452494.87052 | 0.00470 |
| 11030 | 2452495.85265 | 0.00470 |
| 11031 | 2452495.99358 | 0.00470 |
| 11036 | 2452496.69711 | 0.00470 |
| 11037 | 2452496.83754 | 0.00470 |
| 11037 | 2452496.83829 | 0.00470 |
| 11038 | 2452496.97882 | 0.00470 |
| 11039 | 2452497.11997 | 0.00470 |
| 11043 | 2452497.67944 | 0.00470 |
| 11044 | 2452497.81985 | 0.00470 |
| 11050 | 2452498.66128 | 0.00470 |
| 11054 | 2452499.22543 | 0.00470 |
| 11060 | 2452500.06934 | 0.00470 |
| 11061 | 2452500.20953 | 0.00470 |
| 11065 | 2452500.77043 | 0.00470 |
| 11065 | 2452500.77097 | 0.00470 |
| 11066 | 2452500.91146 | 0.00470 |
| 11071 | 2452501.61430 | 0.00470 |
| 11081 | 2452503.01741 | 0.00470 |
| 11082 | 2452503.15864 | 0.00470 |
| 11088 | 2452504.00387 | 0.00470 |
| 11089 | 2452504.14228 | 0.00470 |
| 11108 | 2452506.81640 | 0.00470 |
| 11109 | 2452506.95272 | 0.00470 |
| 11110 | 2452507.09395 | 0.00470 |
| 11116 | 2452507.93727 | 0.00470 |
| 11117 | 2452508.07753 | 0.00470 |
| 11230 | 2452523.98226 | 0.00470 |
| 11235 | 2452524.68604 | 0.00470 |
| 11277 | 2452530.58654 | 0.00470 |
| 11277 | 2452530.58682 | 0.00470 |
| 11284 | 2452531.57057 | 0.00470 |
| 11292 | 2452532.69315 | 0.00470 |
| 11306 | 2452534.66247 | 0.00470 |
| 11336 | 2452538.87909 | 0.00470 |
| 11343 | 2452539.86317 | 0.00470 |
| 11365 | 2452542.95679 | 0.00470 |
| 23838 | 2454296.47818 | 0.00130 |
| 23845 | 2454297.46412 | 0.00130 |
| 23869 | 2454300.83790 | 0.00130 |
| 24081 | 2454330.64265 | 0.00130 |
| 24103 | 2454333.73359 | 0.00130 |
| 24288 | 2454359.74021 | 0.00130 |
| 24266 | 2454356.64910 | 0.00130 |
| 24029 | 2454323.32964 | 0.00130 |
| 24043 | 2454325.30222 | 0.00130 |
| 24044 | 2454325.43949 | 0.00130 |
| 24051 | 2454326.42521 | 0.00130 |
| 24057 | 2454327.26776 | 0.00130 |
| 24079 | 2454330.36156 | 0.00130 |
| 24085 | 2454331.20542 | 0.00130 |
| 24086 | 2454331.34656 | 0.00130 |
| 24087 | 2454331.48360 | 0.00130 |
| 24093 | 2454332.33067 | 0.00130 |
| 24094 | 2454332.46977 | 0.00130 |
| 24101 | 2454333.45237 | 0.00130 |
| 24114 | 2454335.27992 | 0.00130 |
| 24115 | 2454335.42167 | 0.00130 |
| 24121 | 2454336.26490 | 0.00130 |
| 24122 | 2454336.40461 | 0.00130 |
| 24157 | 2454341.32395 | 0.00130 |
| 24164 | 2454342.30913 | 0.00130 |
| 24171 | 2454343.29325 | 0.00130 |
| 24172 | 2454343.43382 | 0.00130 |
| 24179 | 2454344.41706 | 0.00130 |
| 24186 | 2454345.40135 | 0.00130 |
| 24193 | 2454346.38649 | 0.00130 |
| 24228 | 2454351.30712 | 0.00130 |
| 24242 | 2454353.27341 | 0.00130 |
| 24278 | 2454358.33464 | 0.00130 |
| 24285 | 2454359.31780 | 0.00130 |
| 24313 | 2454363.25394 | 0.00130 |
| 26328 | 2454646.51335 | 0.00480 |
| 26498 | 2454670.41927 | 0.00480 |
| 34100 | 2455739.01848 | 0.00190 |
| 34107 | 2455740.00072 | 0.00190 |
| 34121 | 2455741.97212 | 0.00190 |
| 34122 | 2455742.10940 | 0.00190 |
| 34105 | 2455739.71937 | 0.00190 |
| 34362 | 2455775.84363 | 0.00190 |
| 34304 | 2455767.69155 | 0.00190 |
| 34326 | 2455770.78386 | 0.00190 |
| 34468 | 2455790.74504 | 0.00190 |
| 34174 | 2455749.42039 | 0.00190 |
| 34175 | 2455749.55986 | 0.00190 |
| 34189 | 2455751.52421 | 0.00190 |
| 34196 | 2455752.51110 | 0.00190 |
| 34203 | 2455753.49393 | 0.00190 |
| 34217 | 2455755.46037 | 0.00190 |
| 34224 | 2455756.44502 | 0.00190 |
| 34225 | 2455756.58561 | 0.00190 |
| 34196 | 2455752.51121 | 0.00190 |
| 34695 | 2455822.65156 | 0.00190 |
| 34724 | 2455826.72833 | 0.00190 |
| 34943 | 2455857.51562 | 0.00190 |
| 34956 | 2455859.34154 | 0.00190 |

| Spin cycle | Spin dip time (HJD) | Assigned uncertainty (d) |
|---|---|---|
| 34922 | 2455854.56295 | 0.00190 |
| 34951 | 2455858.63794 | 0.00190 |
| 34979 | 2455862.57258 | 0.00190 |
| 36596 | 2456089.86108 | 0.00240 |
| 36703 | 2456104.90021 | 0.00240 |
| 36710 | 2456105.88315 | 0.00240 |
| 36823 | 2456121.76830 | 0.00240 |
| 36831 | 2456122.89056 | 0.00240 |
| 36845 | 2456124.85722 | 0.00240 |
| 36598 | 2456090.14392 | 0.00240 |
| 36605 | 2456091.12676 | 0.00240 |
| 36606 | 2456091.26657 | 0.00240 |
| 36611 | 2456091.97001 | 0.00240 |
| 36668 | 2456099.98093 | 0.00240 |
| 36711 | 2456106.02548 | 0.00240 |
| 36712 | 2456106.16508 | 0.00240 |
| 36725 | 2456107.99490 | 0.00240 |
| 36732 | 2456108.98033 | 0.00240 |
| 36733 | 2456109.11594 | 0.00240 |
| 36739 | 2456109.95994 | 0.00240 |
| 36740 | 2456110.09998 | 0.00240 |
| 36899 | 2456132.44800 | 0.00240 |
| 36913 | 2456134.41534 | 0.00240 |
| 39397 | 2456483.55915 | 0.00180 |
| 39404 | 2456484.54363 | 0.00180 |
| 39716 | 2456528.39909 | 0.00180 |
| 39333 | 2456474.56417 | 0.00180 |
| 39362 | 2456478.63799 | 0.00180 |
| 39434 | 2456488.76145 | 0.00180 |
| 39221 | 2456458.82154 | 0.00180 |
| 39399 | 2456483.84141 | 0.00180 |
| 39442 | 2456489.88278 | 0.00180 |
| 39718 | 2456528.68143 | 0.00180 |
| 39725 | 2456529.66300 | 0.00180 |
| 39732 | 2456530.64621 | 0.00180 |
| 39405 | 2456484.68479 | 0.00180 |
| 39406 | 2456484.82454 | 0.00180 |
| 39412 | 2456485.66880 | 0.00180 |
| 39419 | 2456486.65396 | 0.00180 |
| 39420 | 2456486.79372 | 0.00180 |
| 39427 | 2456487.77532 | 0.00180 |
| 39696 | 2456525.58733 | 0.00180 |
| 39839 | 2456545.68871 | 0.00180 |
| 39846 | 2456546.67362 | 0.00180 |

**Table 3. WD spin cycle numbers, spin dip times and assigned uncertainties.**